\documentclass[letterpaper, 12pt]{article}[2000/05/19]
\usepackage[english]{babel}
\usepackage{amsfonts,amsmath,amssymb,amsthm,latexsym,amscd,mathrsfs}
\usepackage{ifthen,cite}
\usepackage[bookmarksnumbered=true]{hyperref}
\usepackage{times,fancyhdr}

\usepackage{amsmath}
\numberwithin{equation}{section}

\usepackage{graphicx}

\hypersetup{pdfpagetransition={Split}}

\newcommand{\evenhead}{Author \ name}
\newcommand{\oddhead}{Article \ name}
\newcommand{\theArticleName}{Article \ name}

\newcommand{\FirstPageHeading}[1]{\thispagestyle{empty}%
\noindent\raisebox{0pt}[0pt][0pt]{\makebox[\textwidth]{\protect\footnotesize \sf }}\par}

\newcommand{\ArticleName}[1]{\renewcommand{\theArticleName}{#1}\vspace{-2mm}\par\noindent {\LARGE\bf  #1\par}}
\newcommand{\Author}[1]{\vspace{5mm}\par\noindent {\Large  #1\par} \par\vspace{2mm}\par}
\newcommand{\Address}[1]{\vspace{2mm}\par\noindent {\it #1} \par}
\newcommand{\Email}[1]{\ifthenelse{\equal{#1}{}}{}{\par\noindent {\rm E-mail: }{\it  #1} \par}}
\newcommand{\URLaddress}[1]{\ifthenelse{\equal{#1}{}}{}{\par\noindent {\rm URL: }{\tt  #1} \par}}
\newcommand{\EmailD}[1]{\ifthenelse{\equal{#1}{}}{}{\par\noindent {$\phantom{\dag}$~\rm E-mail: }{\it  #1} \par}}
\newcommand{\URLaddressD}[1]{\ifthenelse{\equal{#1}{}}{}{\par\noindent {$\phantom{\dag}$~\rm URL: }{\tt  #1} \par}}

\newcommand{\Abstract}[1]{\vspace{6mm}\par\noindent\hspace*{10mm}
\parbox{140mm}{\small {\bf Abstract.} #1}\par}
\newcommand{\Keywords}[1]{\vspace{3mm}\par\noindent\hspace*{10mm}
\parbox{140mm}{\small {\bf Key words:} \rm #1}\par}
\newcommand{\Classification}[1]{\vspace{3mm}\par\noindent\hspace*{10mm}
\parbox{140mm}{\small {\it 2020 Mathematics Subject Classification:} \rm #1}\vspace{3mm}\par}
\newcommand{\ShortArticleName}[1]{\renewcommand{\oddhead}{#1}}
\newcommand{\AuthorNameForHeading}[1]{\renewcommand{\evenhead}{#1}}

\long\def\@makecaption#1#2{
  \sbox\@tempboxa{\small \textbf{#1.}\ \ #2}%
  \ifdim \wd\@tempboxa >\hsize
    {\small \textbf{#1.}\ \ #2}\par \else
    \global \@minipagefalse
    \hb@xt@\hsize{\hfil\box\@tempboxa\hfil}%
  \fi \vskip\belowcaptionskip}

\def\numberwithin#1#2{\@ifundefined{c@#1}{\@nocounterr{#1}}{%
  \@ifundefined{c@#2}{\@nocnterr{#2}}{%
  \@addtoreset{#1}{#2}%
  \toks@\@xp\@xp\@xp{\csname the#1\endcsname}%
  \@xp\xdef\csname the#1\endcsname
    {\@xp\@nx\csname the#2\endcsname.\the\toks@}}}}
\def\E^#1{{\buildrel #1 \over\vee}}
\newtheorem*{theorem*}{Theorem}
\newtheorem*{criterion*}{Criterion}

{\theoremstyle{definition}

}

\usepackage{tikz}

\begin{document}

\FirstPageHeading{V.I. Gerasimenko, I.V. Gapyak}

\ShortArticleName{Non-perturbative solutions}

\AuthorNameForHeading{V.I. Gerasimenko, I.V. Gapyak}

\ArticleName{\textcolor{blue!55!black}{Non-perturbative solutions of hierarchies\\ of evolution equations for colliding particles}}

\Author{V.I. Gerasimenko$^\ast$\footnote{E-mail: \emph{gerasym@imath.kiev.ua}}
        and I.V. Gapyak$^\ast$$^\ast$\footnote{E-mail: \emph{gapjak@ukr.net}}}

\Address{\hspace*{2mm}$^\ast$Institute of Mathematics of NAS of Ukra\"{\i}ne,\\
    \hspace*{4mm}3, Tereshchenkivs'ka Str.,\\
    \hspace*{4mm}01601, Ky\"{\i}v-4, Ukra\"{\i}ne}

\Address{$^\ast$$^\ast$Taras Shevchenko National University of Ky\"{\i}v,\\
    \hspace*{4mm}Department of Mechanics and Mathematics,\\
    \hspace*{4mm}2, Academician Glushkov Av.,\\
    \hspace*{4mm}03187, Ky\"{\i}v, Ukra\"{\i}ne}

\bigskip

\Abstract{The article deals with the challenge of the construction of solutions to hierarchies of
fundamental evolution equations for many colliding particles. The method of cluster expansions of
the groups of operators of the Liouville equations for observables and a state is used to establish
the generating operators of expansions representing solutions of the Cauchy problems of the BBGKY
hierarchy (Bogolyubov--Born--Green--Kirkwood--Yvon) as well as of the dual BBGKY hierarchy, respectively.}

\bigskip

\bigskip
\Classification{35C10; 35Q20; 82C05; 82D05; 70F45}

\makeatletter
\renewcommand{\@evenhead}{
\hspace*{-3pt}\raisebox{-7pt}[\headheight][0pt]{\vbox{\hbox to \textwidth {\thepage \hfil \evenhead}\vskip4pt \hrule}}}
\renewcommand{\@oddhead}{
\hspace*{-3pt}\raisebox{-7pt}[\headheight][0pt]{\vbox{\hbox to \textwidth {\oddhead \hfil \thepage}\vskip4pt\hrule}}}
\renewcommand{\@evenfoot}{}
\renewcommand{\@oddfoot}{}
\makeatother

\newpage
\vphantom{math}

\protect\textcolor{blue!55!black}{\tableofcontents}


\textcolor{blue!55!black}{\section{Introduction}}

At the present time, a number of works have appeared, in which possible approaches to the problem of
the description of the evolution of many colliding particles \cite{GG21}--\cite{AP24} are discussed
as well as of many quantum particles \cite{BPS}; in particular, this is related to the challenge of
the rigorous derivation of kinetic equations from the underlying hierarchies of fundamental evolution
equations \cite{B1}--\cite{GG23}.

The traditional approach to describing the evolution of both finitely and infinitely many classical
particles is based on the description of the evolution of all possible states by means of the reduced
distribution functions governed by the BBGKY hierarchy (Bogolyubov--Born--Green--Kirkwood--Yvon),
which for finitely many particles is equivalent to the Liouville equation for the probability
distribution function \cite{CGP97}--\cite{PG90}. As is well known, the advantages of this approach
consist of the possibility of rigorously describing the evolution of infinitely many particles whose
collective behavior exhibits thermodynamic (statistical) features, namely, the existence of an
equilibrium state in such a system as well as the kinetic or hydrodynamic behavior in corresponding
scaling approximations \cite{CGP97}--\cite{R69}.

From the moment the BBGKY hierarchy was formulated in 1946 until the last decade, the solution to
such a hierarchy of evolution equations has been represented in the form of an iteration series,
i.e., expansion into a series constructed by perturbation theory methods. In particular, this
representation of the solution is applied for the derivation of kinetic equations \cite{L75}--\cite{CGP97},
the generally accepted method of derivation of which is the construction of the scaling asymptotics of the
solution to the BBGKY hierarchy.

Let us make some comments concerning the existence of solutions to the Cauchy problem of the BBGKY
hierarchy for initial data from various function spaces. In the spaces of sequences of integrable
functions, the existence and uniqueness of a global-in-time solution of the BBGKY hierarchy were
proved in the papers \cite{PG83},\cite{GR02},\cite{GerRS} (see also book \cite{CGP97}).

To describe the evolution of a state of infinitely many classical particles, the suitable functional
space is the space of sequences of functions bounded with respect to the configuration variables and
decreasing with respect to the momentum ones; in particular, the equilibrium distribution functions
belong to this space \cite{R69}. For the solution extension from the space of sequences of integrable
functions to this space, the method of the thermodynamic limit was developed \cite{CGP97},\cite{GP85}.

As mentioned above, for a solution representation of the Cauchy problem of the BBGKY hierarchy for hard
spheres is widely used in the representation as a series of perturbation theory (an iteration series over
the evolution of the states of selected groups of particles) \cite{B1}--\cite{CIP},\cite{GP85}. In this
form, the solution is applied to construct its Boltzmann--Grad asymptotics, which is governed by the
Boltzmann kinetic equation. The justification of a solution represented as an iteration series for hard
spheres is based on giving a rigorous mathematical meaning to every term of the iteration series and on
the proof of its convergence. The main difficulty in this problem is that the phase trajectories of
particles for a system with a singular interaction potential are defined almost everywhere in the phase
space, and initial distribution functions in the iteration series are concentrated on lower-dimensional
manifolds. It is necessary to ensure that the trajectories are defined on these manifolds. This problem
was completely solved in the papers \cite{PG90},\cite{GP90}.

In the case of infinitely many hard spheres, a local-in-time solution \cite{GP85} of the Cauchy problem
of the BBGKY hierarchy is represented by an iteration series for arbitrary initial data from the space
of sequences of functions bounded with respect to configuration variables, and for initial data close
to the equilibrium state, it is a global-in-time solution \cite{PG90}. For such initial data in a
one-dimensional space for a hard sphere system, the existence of a global in-time solution was proved
in the paper \cite{Ger92}.

In recent papers \cite{GG18}--\cite{GG11} an approach has been developed to a rigorous derivation of
the kinetic equations for many particles interacting as hard spheres based on the description of the
evolution of observables governed by the dual BBGKY hierarchy for the so-called reduced observables
\cite{BGer}--\cite{D16}. This approach involves constructing the asymptotic behavior of the solution
to the dual BBGKY hierarchy for reduced observables. In a similar approach, the description of the
kinetic evolution in the mean-field scaling limit within the framework of observables for quantum
many-particle systems was developed in papers \cite{G11}--\cite{G13}.

This article argues the structure of expansions that represent a non-perturbative solution of the
Cauchy problem of the dual BBGKY hierarchy for observables of many hard spheres as well as expansions
into series that represent a non-perturbative solution of the BBGKY hierarchy for their state.

\textcolor{blue!55!black}{\section{The dual BBGKY hierarchy for colliding particles}}

At the beginning, we introduce some preliminary concepts accepted for describing many colliding
particles.

Within the framework of a non-fixed, i.e., arbitrary but finite average number of particles
(non-equilibrium grand canonical ensemble), the observables and a state of a hard sphere system
are described by the sequences of functions $A(t)=(A_0,A_{1}(t,x_1),\ldots,A_{n}(t,x_1,\ldots,x_n),\ldots)$
at instant $t\in\mathbb{R}$ and of a sequence $D(0)=(D_0,D_{1}^0(x_1),\ldots,D_{n}^0(x_1,\ldots,x_n),\ldots)$
of the probability distribution functions at initial moment defined on the phase spaces of the corresponding
number of hard spheres, i.e., $x_i\equiv(q_i,p_i)\in\mathbb{R}^{3}\times\mathbb{R}^{3}$ is phase coordinates
that characterize a center of the $ith$ hard sphere in the space $\mathbb{R}^{3}$ and its momentum \cite{CGP97}.
For configurations of a system of identical particles of a unit mass interacting as hard spheres with
a diameter of $\sigma>0$ the following inequalities are satisfied: $|q_i-q_j|\geq\sigma,$ $i\neq j\geq1$,
i.e. the set $\mathbb{W}_n\equiv\big\{(q_1,\ldots,q_n)\in\mathbb{R}^{3n}\big||q_i-q_j|<\sigma$ for at
least one pair $(i,j):\,i\neq j\in(1,\ldots,n)\big\}$, $n>1$, is the set of forbidden configurations.

Let $\mathcal{C}_{\gamma}$ be the space of sequences $b=(b_0,b_1,\ldots,b_n,\ldots)$ of the bounded continuous
functions $b_n(x_1,\ldots,x_n)$ that are symmetric with respect to permutations of the arguments $x_1,\ldots,x_n$,
equal to zero on the set of forbidden configurations $\mathbb{W}_n$ and equipped with the norm:
\begin{eqnarray*}
   &&\|b\|_{\mathcal{C}_{\gamma}}=\max_{n\geq 0}\,\frac{\gamma^{n}}{n!}\,\|b\|_{\mathcal{C}_{n}}=
            \max_{n\geq 0}\,\frac{\gamma^{n}}{n!}\,\sup_{x_{1},\ldots,x_{n}}|b_n(x_{1},\ldots,x_{n})|,
\end{eqnarray*}
where $0<\gamma<1$. We also introduce the space $L^{1}_{\alpha}=\oplus^{\infty}_{n=0}\alpha^n L^{1}_{n}$
of sequences $f=(f_0,f_1,\ldots,f_n,\ldots)$ of integrable functions $f_n(x_1,\ldots,x_n)$ that are
symmetric with respect to permutations of the arguments $x_1,\ldots,x_n$, equal to zero on the set
$\mathbb{W}_n$ and equipped with the norm:
\begin{eqnarray*}
   &&\|f\|_{L^{1}_{\alpha}}=\sum_{n=0}^\infty \alpha^n
        \int\limits_{(\mathbb{R}^{3}\times\mathbb{R}^{3})^n}|f_n(x_1,\ldots,x_n)|dx_1\ldots dx_n,
\end{eqnarray*}
where $\alpha>1$ is a real number.

A mean value functional of observables is represented by the following series expansion \cite{PG90}:
\begin{eqnarray}\label{averageD}
   &&\hskip-8mm \langle A\rangle(t)=(I,D(0))^{-1}(A(t),D(0)),
\end{eqnarray}
where such an abbreviation is used:
\begin{eqnarray*}
    &&\hskip-8mm (A(t),D(0))=\sum_{n=0}^\infty\,\frac{1}{n!}
         \int\limits_{(\mathbb{R}^{3}\times\mathbb{R}^{3})^n}
         A_{n}(t,x_1,\ldots,x_n)D_{n}^0(x_1,\ldots,x_n)dx_1\ldots dx_n,
\end{eqnarray*}
and the coefficient $(I,D(0))$ is a normalizing factor, i.e., $I\doteq (1,\ldots,1,\ldots)$.

If $A(t)\in\mathcal{C}_{\gamma}$ and $D(0)\in L^{1}_{\alpha}$ mean value functional (\ref{averageD})
exists and determines the duality between observables and a state of many hard spheres.

The evolution of the observables is described by the Cauchy problem for the sequence of the weak
formulation of the Liouville equations for many hard spheres \cite{GG21}.
On the space $\mathcal{C}_{\gamma}$ the non-perturbative solution of the abstract Cauchy problem
to the Liouville equations
\begin{eqnarray}\label{Le}
   &&A(t)=S(t)A(0)
\end{eqnarray}
is determined by the following group of operators \cite{PG90}:
\begin{eqnarray}\label{Sspher}
  &&\hskip-8mm(S(t)b)_{n}(x_{1},\ldots,x_{n})\equiv S_{n}(t,1,\ldots,n)b_{n}(x_{1},\ldots,x_{n})\doteq \\
  &&    \begin{cases}
         b_{n}(X_{1}(t,x_{1},\ldots,x_{n}),\ldots,X_{n}(t,x_{1},\ldots,x_{n})),\hskip+5mm \mathrm{if}\,
         (x_{1},\ldots,x_{n})\in(\mathbb{R}^{3n}\times(\mathbb{R}^{3n}\setminus\mathbb{W}_{n})),\nonumber\\
         0, \hskip+78mm \mathrm{if}\, (q_{1},\ldots,q_{n})\in\mathbb{W}_{n},
         \end{cases}
\end{eqnarray}
where for $t\in\mathbb{R}$ the function $X_{i}(t)$ is a phase trajectory of $i-th$ particle constructed
in book \cite{CGP97} and the set $\mathbb{M}_{n}^0$ consists of the phase space points which are specified
such initial data that during the evolution generate multiple collisions, i.e. collisions of more than two
particles, more than one two-particle collision at the same instant and infinite number of collisions within
a finite time interval.

On the space $\mathcal{C}_{\gamma}$, one-parameter mapping (\ref{Sspher}) is an isometric $\ast$-weak
continuous group of operators, i.e., it is a $C_{0}^{\ast}$-group. The infinitesimal generator
$\mathcal{L}=\oplus_{n=0}^\infty\mathcal{L}_{n}$ of the group of operators (\ref{Sspher}) has the structure:
$\mathcal{L}_{n}=\sum_{j=1}^{n}\mathcal{L}(j)+\sum_{j_{1}<j_{2}=1}^{n}\mathcal{L}_{\mathrm{int}}(j_{1},j_{2}),$
where the operator $\mathcal{L}(j)\doteq\langle p_j,\frac{\partial}{\partial q_j}\rangle$ defined on the set
$C_{n,0}$ of continuously differentiable functions with compact supports is the Liouville operator of free
motion of the $j$ hard sphere and in the case of $t\geq0$ the operator describing collisions
$\mathcal{L}_{\mathrm{int}}(j_1,j_{2})$ is defined by the formula  \cite{PG90}
\begin{eqnarray}\label{int}
   &&\hskip-18mm\mathcal{L}_{\mathrm{int}}(j_1,j_{2})b_n\doteq
     \sigma^{2}\int\limits_{\mathbb{S}_+^2}d\eta\langle\eta,(p_{j_1}-p_{j_2})\rangle
     \big(b_n(x_1,\ldots,q_{j_1},p_{j_1}^*,\ldots,q_{j_2},p_{j_2}^*,\ldots,x_n)-\\
   &&-b_n(x_1,\ldots,x_n)\big)\delta(q_{j_1}-q_{j_2}+\sigma\eta).\nonumber
\end{eqnarray}
In this formula (\ref{int}) the symbol $\langle \cdot,\cdot \rangle$ denotes a scalar product, $\delta$ is the Dirac measure,
$\mathbb{S}_{+}^{2}\doteq\{\eta\in\mathbb{R}^{3}\big|\left|\eta\right|=1,\langle\eta,(p_{j_{1}}-p_{j_{2}})\rangle>0\}$
and the post-collision momenta $p_{j_{1}}^*,p_{j_{2}}^*$ are defined by the equalities
\begin{eqnarray}\label{momenta}
     &&p_{j_{1}}^*\doteq p_{j_{1}}-\eta\left\langle\eta,\left(p_{j_{1}}-p_{j_{2}}\right)\right\rangle, \\
     &&p_{j_{2}}^*\doteq p_{j_{2}}+\eta\left\langle\eta,\left(p_{j_{1}}-p_{j_{2}}\right)\right\rangle \nonumber.
\end{eqnarray}
For $t<0$ operator (\ref{int}) is defined by the corresponding expression \cite{GG21}.

The motivation for describing the evolution of many-particle systems in terms of the so-called reduced
observables is associated with possible equivalent representations of the functional of the mean value
(mathematical expectation) of the observables.

To formulate the representation of mean value functional (\ref{averageD}) in terms of sequences of
reduced observables and reduced distribution functions on sequences of bounded continuous functions,
we introduce an analog of the creation operator
\begin{eqnarray}\label{a+}
   &&(\mathfrak{a}^{+}b)_{s}(x_{1},\ldots,x_{s})\doteq\sum_{j=1}^s\,b_{s-1}((x_{1},\ldots,x_{s})\setminus (x_j)),
\end{eqnarray}
and on sequences of integrable functions we introduce an adjoint operator to operator (\ref{a+}) in
the sense of mean value functional (\ref{averageD}) which is an analogue of the annihilation operator
\begin{eqnarray}\label{a}
    &&(\mathfrak{a}f)_{n}(x_1,\ldots,x_n)=
      \int\limits_{\mathbb{R}^3\times\mathbb{R}^3}f_{n+1}(x_1,\ldots,x_n,x_{n+1})dx_{n+1}.
\end{eqnarray}
Then as a consequence of the validity of equalities:
$$(b,f)=(e^{\mathfrak{a^{+}}}e^{-\mathfrak{a^{+}}}b,f)=(e^{-\mathfrak{a^{+}}}b,e^{\mathfrak{a}}f),$$
for mean value functional (\ref{averageD}) the following representation holds:
\begin{eqnarray}\label{B(t)}
   &&\langle A\rangle(t)=(I,D(0))^{-1}(A(t),D(0))=(B(t),F(0)),
\end{eqnarray}
where a sequence of the reduced observables is defined by the formula
\begin{eqnarray}\label{ro}
   && B(t)=e^{-\mathfrak{a^{+}}}A(t),
\end{eqnarray}
and a sequence of so-called reduced distribution functions is defined as follows (known as the
non-equilibrium grand canonical ensemble \cite{PG83})
\begin{eqnarray*}\label{rdf}
   && F(0)=(I,D(0))^{-1}e^{\mathfrak{a}}D(0).
\end{eqnarray*}

Thus, according to the definition of the operator $e^{-\mathfrak{a^{+}}}$, the sequence of reduced
observables (\ref{ro}) in component-wise form is represented by the expansions:
\begin{eqnarray}\label{moo}
  &&\hskip-12mm B_s(t,x_1,\ldots,x_s)=\sum_{n=0}^s\,\frac{(-1)^n}{n!}\sum_{j_1\neq\ldots\neq j_{n}=1}^s
        A_{s-n}\big(t,(x_1,\ldots,x_s)\setminus (x_{j_1},\ldots,x_{j_{n}})), \quad s\geq1.
\end{eqnarray}

Note that the evolution of many hard spheres is traditionally described in terms of the evolution
of states governed by the BBGKY hierarchy for reduced distribution functions (\ref{rdf}). An equivalent
approach to describing evolution is based on reduced observables governed by a dual hierarchy.

The evolution of sequence of reduced observables (\ref{ro}) of many hard spheres is determined by
the Cauchy problem of the following abstract hierarchy of evolution equations \cite{GG18},\cite{BGer}:
\begin{eqnarray}\label{dh}
   &&\hskip-8mm \frac{d}{dt}B(t)=\mathcal{L}B(t)+\big[\mathcal{L},\mathfrak{a}^+\big]B(t),\\
   \nonumber\\
   \label{dhi}
   &&\hskip-8mm B(t)|_{t=0}=B(0),
\end{eqnarray}
where the operator $\mathcal{L}$ is generator (\ref{int}) of the group of operators (\ref{Sspher}) for
hard spheres, the symbol $\big[ \cdot,\cdot \big]$ denotes the commutator of operators, which in the
abstract evolution equation (\ref{dh}) has the following component-wise form:
\begin{eqnarray*}
   &&\hskip-8mm (\big[\mathcal{L},\mathfrak{a}^+\big]b)_{s}(x_1,\ldots,x_s)=
       \sum_{j_1\neq j_{2}=1}^s\mathcal{L}_{\mathrm{int}}(j_1,j_{2})b_{s-1}(t,(x_1,\ldots,x_s)\setminus x_{j_1}),
       \quad s\geq2.
\end{eqnarray*}
In a component-wise form, the hierarchy of evolution equations (\ref{dh}) for hard spheres, in fact,
is a sequence of recurrence evolution equations (in literature, it is known as the dual BBGKY hierarchy \cite{GG21}).
We adduce the simplest examples of recurrent evolution equations (\ref{dh}):
\begin{eqnarray*}
   &&\hskip-8mm \frac{\partial}{\partial t}B_{1}(t,x_1)=\mathcal{L}(1)B_{1}(t,x_1),\\
   &&\hskip-8mm \frac{\partial}{\partial t}B_{2}(t,x_1,x_2)=
      \big(\sum_{i=1}^{2}\mathcal{L}(j)+\mathcal{L}_{\mathrm{int}}(1,2)\big)B_{2}(t,x_1,x_2)+
      \mathcal{L}_{\mathrm{int}}(1,2)\big(B_{1}(t,x_1)+B_{1}(t,x_2)\big),\\
   &&\vdots
\end{eqnarray*}
where the generators of these evolution equations are defined by formulas (\ref{int}).

\textcolor{blue!55!black}{\section{A non-perturbative solution of the dual BBGKY hierarchy}}

A non-perturbative solution of the Cauchy problem of the dual BBGKY hierarchy (\ref{dh}),(\ref{dhi}) for hard
spheres is a sequence of reduced observables represented by the following expansions \cite{GR02},\cite{GR03}:
\begin{eqnarray}\label{sed}
   &&\hskip-12mm B_{s}(t,x_1,\ldots,x_s)=(e^{\mathfrak{a}^{+}}\mathfrak{A}(t)B(0))_s(x_1,\ldots,x_s)=\\
   &&\hskip-8mm\sum_{n=0}^s\,\frac{1}{n!}\sum_{j_1\neq\ldots\neq j_{n}=1}^s
      \mathfrak{A}_{1+n}(t,\{(1,\ldots,s)\setminus(j_1,\ldots,j_{n})\},j_1,\ldots,j_{n}\big)\,
      B_{s-n}^{0}(x_1,\ldots\nonumber\\
   && \ldots,x_{j_1-1},x_{j_1+1},\ldots,x_{j_n-1},x_{j_n+1},\ldots,x_s),
      \quad s\geq1,\nonumber
\end{eqnarray}
where the mappings $\mathfrak{A}_{1+n}(t),\,n\geq0,$ are the generating operators which are represented
as cumulant expansions with respect of the groups of operators (\ref{Sspher}).

The simplest examples of reduced observables (\ref{sed}) are given by the expansions:
\begin{eqnarray*}
   &&B_{1}(t,x_1)=\mathfrak{A}_{1}(t,1)B_{1}^{0}(x_1),\\
   &&B_{2}(t,x_1,x_2)=\mathfrak{A}_{1}(t,\{1,2\})B_{2}^{0}(x_1,x_2)+
      \mathfrak{A}_{2}(t,1,2)(B_{1}^{0}(x_1)+B_{1}^{0}(x_2)),\\
   &&\vdots
\end{eqnarray*}

To determine the generating operators of expansions of reduced observables (\ref{sed}),
we will introduce the notion of dual cluster expansions of the groups of operators (\ref{Sspher})
in terms of operators interpreted as their cumulants. For this end on sequences of one-parametric
mappings $\mathfrak{u}(t)=(0,\mathfrak{u}_1(t),\ldots,\mathfrak{u}_n(t),\ldots)$ we introduce the
following notion of $\star$-product \cite{R69}:
\begin{eqnarray}\label{Product}
    &&\hskip-8mm(\mathfrak{u}(t)\star\widetilde{\mathfrak{u}}(t))_{s}(1,\ldots,s)\doteq
        \sum\limits_{Y\subset (1,\ldots,s)}\,\mathfrak{u}_{|Y|}(t,Y)
        \,\widetilde{\mathfrak{u}}_{s-|Y|}(t,(1,\ldots,s)\setminus Y),
\end{eqnarray}
where $\sum_{Y\subset (1,\ldots,s)}$ is the sum over all subsets $Y$ of the set $(1,\ldots,s)$.

Using the definition of the $\star$-product (\ref{Product}), the dual cluster expansions of groups of
operators (\ref{Sspher}) are represented by the mapping ${\mathbb E}\mathrm{xp}_{\star}$ in the form
\begin{eqnarray*}\label{DtoGcircledStar}
    && S(t)={\mathbb E}\mathrm{xp}_{\star}\,\mathfrak{A}(t)=\mathbb{I}+
      \sum\limits_{n=1}^{\infty}\frac{1}{n!}\mathfrak{A}(t)^{\star n},
\end{eqnarray*}
where $S(t)=(0,S_1(t,1),\ldots,S_n(t,1,\ldots,n),\ldots)$ and $\mathbb{I}=(1,0,\ldots,0,\ldots)$.
In component-wise form, the dual cluster expansions are represented by the following
recursive relations:
\begin{eqnarray}\label{cexd}
   &&\hskip-8mm S_{s}(t,(1,\ldots,s)\setminus(j_1,\ldots,j_{n}),j_1,\ldots,j_{n})=\\
   &&\sum\limits_{\mathrm{P}:\,(\{(1,\ldots,s)\setminus(j_1,\ldots,j_{n})\},\,j_1,\ldots,j_{n})=
       \bigcup_i X_i}\,\prod\limits_{X_i\subset\mathrm{P}}\mathfrak{A}_{|X_i|}(t,X_i),\quad n\geq 0,\nonumber
\end{eqnarray}
where the set consisting of one element of indices $(1,\ldots,s)\setminus(j_1,\ldots,j_{n})$ we denoted by
the symbol $\{(1,\ldots,s)\setminus(j_1,\ldots,j_{n})\}$ and the symbol ${\sum}_\mathrm{P}$ means the sum
over all possible partitions $\mathrm{P}$ of the set $(\{(1,\ldots,s)\setminus(j_1,\ldots,j_{n})\},\,j_1,\ldots,j_{n})$
into $|\mathrm{P}|$ nonempty mutually disjoint subsets $X_i\subset(1,\ldots,s)$.

The solutions of recursive relations (\ref{cexd}) are represented by the inverse mapping
${\mathbb L}\mathrm{n}_{\ast}$ in the form of the cumulant expansion
\begin{eqnarray*}
    && \mathfrak{A}(t)={\mathbb L}\mathrm{n}_{\star}(\mathbb{I}+S(t))=
         \sum\limits_{n=1}^{\infty}\frac{(-1)^{n-1}}{n}S(t)^{\star n}.
\end{eqnarray*}
Then the $(1+n)th$-order dual cumulant of groups of operators (\ref{Sspher}) is defined by
the following expansion:
\begin{eqnarray}\label{dcumulant}
    &&\hskip-12mm \mathfrak{A}_{1+n}(t,\{(1,\ldots,s)\setminus(j_1,\ldots,j_{n})\},j_1,\ldots,j_{n})\doteq\\
    &&\hskip-5mm\sum\limits_{\mathrm{P}:\,(\{(1,\ldots,s)\setminus(j_1,\ldots,j_{n})\},j_1,\ldots,j_{n})={\bigcup}_i X_i}
       (-1)^{\mathrm{|P|}-1}({\mathrm{|P|}-1})!\prod_{X_i\subset\mathrm{P}}
       S_{|\theta(X_i)|}(t,\theta(X_i)),\nonumber
\end{eqnarray}
where the above notation is used and the declusterization mapping $\theta$ is defined by
the formula: $\theta(\{(1,\ldots,$ $s)\setminus(j_1,\ldots,j_{n})\})=(1,\ldots,s)\setminus(j_1,\ldots,j_{n})$.
The dual cumulants (\ref{dcumulant}) of groups of operators (\ref{Sspher}) of the first two
orders have the form:
\begin{eqnarray*}
    &&\hskip-8mm\mathfrak{A}_{1}(t,\{1,\ldots,s\})=S_{s}(t,1,\ldots,s),\\
    &&\hskip-8mm\mathfrak{A}_{1+1}(t,\{(1,\ldots,s)\setminus(j)\},j)=
       S_{s}(t,1,\ldots,s)-S_{s-1}(t,(1,\ldots,s)\setminus(j))S_{1}(t,j),\\
   &&\vdots
\end{eqnarray*}

If $b_{s}\in\mathcal{C}_{s}$, then for $(1+n)th$-order cumulant (\ref{dcumulant}) of the groups
of operators (\ref{Sspher}) the following estimate is valid:
\begin{eqnarray}\label{estd}
   &&\hskip-8mm \big\|\mathfrak{A}_{1+n}(t)b_{s}\big\|_{\mathcal{C}_{s}}
      \leq \sum_{\mathrm{P}:\,(\{(1,\ldots,s)\setminus(j_1,\ldots,j_{n})\},j_1,\ldots,j_{n})={\bigcup}_i X_i}
      (|\mathrm{P}|-1)!\big\|b_{s}\big\|_{\mathcal{C}_{s}}\leq \\
   &&\sum\limits_{k=1}^{n+1}\mathrm{s}(n+1,k)(k-1)!\big\|b_{s}\big\|_{\mathcal{C}_{s}}
      \leq n!e^{n+2}\big\|b_{s}\big\|_{\mathcal{C}_{s}},\nonumber
\end{eqnarray}
where $\mathrm{s}(n+1,k)$ are the Stirling numbers of the second kind. Then according to this estimate
(\ref{estd}) for the generating operators of expansions (\ref{sed}) provided that $\gamma<e^{-1}$
the inequality valid
\begin{eqnarray}\label{esd}
    &&\hskip-8mm \big\|B(t)\big\|_{C_{\gamma}}\leq e^2(1-\gamma e)^{-1}\big\|B(0)\big\|_{C_{\gamma}}.
\end{eqnarray}

In fact, the following criterion holds.

\begin{criterion*}
A solution of the Cauchy problem of the dual BBGKY hierarchy (\ref{dh}),(\ref{dhi}) is represented by
expansions (\ref{sed}) if and only if the generating operators of expansions (\ref{sed}) are solutions
of cluster expansions (\ref{cexd}) of the groups of operators (\ref{Sspher}) of the Liouville equations
for hard spheres.
\end{criterion*}

The necessity condition means that cluster expansions (\ref{cexd}) take place for groups of operators
(\ref{Sspher}). These recurrence relations are derived from definition (\ref{moo}) of reduced observables,
provided that they are represented as expansions (\ref{sed}) for the solution of the Cauchy problem
of the dual BBGKY hierarchy (\ref{dh}),(\ref{dhi}).

The sufficient condition means that the infinitesimal generator of one-parameter mapping (\ref{sed})
coincides with the generator of the sequence of recurrence evolution equations (\ref{dh}).
Indeed, in the space $C_{\gamma}$ the following existence theorem is true \cite{GG18}.
\begin{theorem*}
A non-perturbative solution of the Cauchy problem (\ref{dh}),(\ref{dhi}) is represented by
expansions (\ref{sed}) in which the generating operators are cumulants of the corresponding
order (\ref{dcumulant}) of the groups of operators (\ref{Sspher}):
\begin{eqnarray}\label{sedc}
   &&\hskip-15mm B_{s}(t,x_1,\ldots,x_s)=\sum_{n=0}^s\,\frac{1}{n!}\sum_{j_1\neq\ldots\neq j_{n}=1}^s\,
      \sum\limits_{\mathrm{P}:\,(\{(1,\ldots,s)\setminus(j_1,\ldots,j_{n})\},j_1,\ldots,j_{n})={\bigcup}_i X_i}
       (-1)^{\mathrm{|P|}-1}({\mathrm{|P|}-1})!\times \nonumber \\
   &&\hskip-5mm\prod_{X_i\subset\mathrm{P}}
       S_{|\theta(X_i)|}(t,\theta(X_i))B_{s-n}^{0}(x_1,\ldots,x_{j_1-1},x_{j_1+1},\ldots,x_{j_n-1},x_{j_n+1},\ldots,x_s),
      \quad s\geq1.\nonumber
\end{eqnarray}
Under the condition $\gamma<e^{-1}$ for initial data $B(0)\in C_{\gamma}^0$ of finite sequences
of infinitely differentiable functions with compact supports sequence (\ref{sedc}) is a unique
global-in-time classical solution, and for arbitrary initial data $B(0)\in C_{\gamma}$ is a unique
global-in-time generalized solution.
\end{theorem*}

The proof of this theorem is similar to the proof of the existence theorem for the Liouville equations
for hard spheres \cite{GG18}.

We note that the one component sequences $B^{(1)}(0)=(0,b_{1}(x_1),0,\ldots)$ of reduced observables
correspond to the additive-type observable, and the sequences $B^{(k)}(0)=(0,\ldots,b_{k}(x_1,\ldots,x_k),0,\ldots)$
of reduced observables correspond to the $k$-ary-type observables \cite{BGer}.

If initial data (\ref{dhi}) is specified by the additive-type reduced observable, then the structure
of solution expansion (\ref{sedc}) is simplified and attains the form
\begin{eqnarray*}\label{af}
     &&B_{s}^{(1)}(t,x_1,\ldots,x_s)=\mathfrak{A}_{s}(t,1,\ldots,s)\sum_{j=1}^s b_{1}(x_j), \quad s\geq 1,
\end{eqnarray*}
where the generating operator $\mathfrak{A}_{s}(t)$ is the $sth$-order cumulant (\ref{dcumulant}) of
the groups of operators (\ref{Sspher}). An example of the additive-type observables is a number of particles,
i.e., the sequence $N^{(1)}(0)=(0,1,0,\ldots)$, then
\begin{eqnarray*}
    &&\hskip-8mm N^{(1)}_{s}(t)=\mathfrak{A}_{s}(t,1,\ldots,s)s=\sum\limits_{\mathrm{P}:\,(1,\ldots,s)={\bigcup}_i X_i}
        (-1)^{\mathrm{|P|}-1}({\mathrm{|P|}-1})!\sum_{j=1}^s 1=\\
    &&\sum\limits_{k=1}^s(-1)^{k-1}\mathrm{s}(s,k)(k-1)! s=s\delta_{s,1}=N^{(1)}_{s}(0),
\end{eqnarray*}
where $\mathrm{s}(s,k)$ are the Stirling numbers of the second kind and $\delta_{s,1}$ is a Kronecker symbol.
Consequently, the observable of a number of hard spheres is an integral of motion and, in particular, the
average number of particles is preserving in time.

In the case of initial $k$-ary-type, $k\geq2$, reduced observables solution expansion (\ref{sedc})
takes the form
\begin{eqnarray*}\label{af-k}
     &&\hskip-15mm B_{s}^{(k)}(t)=0, \quad 1\leq s<k,\\
     &&\hskip-15mm B_{s}^{(k)}(t,x_1,\ldots,x_s)=\frac{1}{(s-k)!}\sum_{j_1\neq\ldots\neq j_{s-k}=1}^s
      \mathfrak{A}_{1+s-k}\big(t,\{(1,\ldots,s)\setminus (j_1,\ldots,j_{s-k})\},\nonumber\\
     &&j_1,\ldots,j_{s-k}\big)\,
      b_{k}(x_1,\ldots,x_{j_1-1},x_{j_1+1},\ldots,x_{j_s-k-1},x_{j_s-k+1},\ldots,x_s),\quad s\geq k,\nonumber
\end{eqnarray*}
where the generating operator $\mathfrak{A}_{1+s-k}(t)$ is the $(1+s-k)th$-order cumulant (\ref{dcumulant})
of the groups of operators (\ref{Sspher}).

We emphasize that cluster expansions (\ref{cexd}) of the groups of operators (\ref{Sspher}) underlie
the classification of possible solution representations of the Cauchy problem (\ref{dh}),(\ref{dhi})
of the dual BBGKY hierarchy. Indeed, using cluster expansions (\ref{cexd}) of the groups of operators
(\ref{Sspher}), other solution representations can be constructed.
For example, let us express the cumulants $\mathfrak{A}_{1+n}(t),\,n\geq2,$ of the groups of operators
(\ref{Sspher}) with respect to the $1st$-order and $2nd$-order cumulants. The following equalities are
true:
\begin{eqnarray*}
   &&\hskip-7mm \mathfrak{A}_{1+n}(t,\{(1,\ldots,s)\setminus(j_1,\ldots,j_{n})\},j_1,\ldots,j_{n})=\\
   &&\hskip-5mm \sum_{Y\subset(j_1,\ldots,j_{n}),\, Y\neq \emptyset}
       \mathfrak{A}_{2}(t,\{(1,\ldots,s)\setminus(j_1,\ldots,j_{n})\}, \{Y\})
       \sum_{\mathrm{P}:\,(j_1,\ldots,j_{n})\setminus Y ={\bigcup\limits}_i X_i}(-1)^{|\mathrm{P}|}\,|\mathrm{P}|!\,
      \prod_{i=1}^{|\mathrm{P}|}\mathfrak{A}_{1}(t,\{X_{i}\}),\\
   &&\hskip-7mm n\geq2,
\end{eqnarray*}
where ${\sum\limits}_{Y\subset(j_1,\ldots,j_{n}),\,Y\neq\emptyset}$ is a sum over all nonempty subsets
$Y\subset (j_1,\ldots,j_{n})$. Then, taking into account the identity
\begin{eqnarray}\label{id}
  &&\hskip-21mm \sum_{\mathrm{P}:\,(j_1,\ldots,j_{n})\setminus Y ={\bigcup\limits}_i X_i}
     (-1)^{|\mathrm{P}|}\,|\mathrm{P}|!\,\prod_{i=1}^{|\mathrm{P}|}
     \mathfrak{A}_{1}(t,\{X_{i}\})B^0_{s-n}((x_{1},\ldots,x_{s})\setminus(x_{j_1},\ldots,x_{j_{n}}))=\\
   &&\sum_{\mathrm{P}:\,(j_1,\ldots,j_{n})\setminus Y ={\bigcup\limits}_i X_i}
     (-1)^{|\mathrm{P}|}\,|\mathrm{P}|!\,B^0_{s-n}((x_{1},\ldots,x_{s})\setminus(x_{j_1},\ldots,x_{j_{n}})),\nonumber
\end{eqnarray}
and the equalities
\begin{eqnarray}\label{aq}
  &&{\sum_{\mathrm{P}:\,(j_1,\ldots,j_{n})\setminus Y =
      {\bigcup\limits}_i X_i}}(-1)^{|\mathrm{P}|}\,|\mathrm{P}|!=
      (-1)^{|(j_1,\ldots,j_{n})\setminus Y|},
      \quad Y\subset\,(j_1,\ldots,j_{n}),
\end{eqnarray}
for solution expansions (\ref{sed}) of the dual BBGKY hierarchy we derive the following representation:
\begin{eqnarray*}
    &&\hskip-15mm B _{s}(t,x_1,\ldots,x_s)=\mathfrak{A}_{1}(t,\{1,\ldots,s\})B_{s}^0(x_1,\ldots,x_s)+\\
    && \sum_{n=1}^s\,\frac{1}{n!}\,\sum_{j_1\neq\ldots\neq j_{n}=1}^s\,\,
       \sum\limits_{Y\subset (1,\ldots,s)\setminus(j_1,\ldots,j_{n}),\,Y\neq \emptyset}\,
       (-1)^{|(j_1,\ldots,j_{n})\setminus Y|}\,\times\\
    && \mathfrak{A}_{2}(t,\{j_1,\ldots,j_{n}\},\{Y\})\,B^0_{s-n}((x_{1},\ldots,x_{s})\setminus(x_{j_1},\ldots,x_{j_{n}})),
    \quad s\geq 1,
\end{eqnarray*}
where notations accepted above are used.


\textcolor{blue!55!black}{\section{Representation of reduced observables in the form of iteration expansions}}

An analog of solution expansions of the BBGKY hierarchy constructed in \cite{PG83},\cite{GerRS}
in the case of the dual BBGKY hierarchy (\ref{dh}) is represented by expansions (\ref{sed}) with
the generating operators, which are reduced cumulants of groups of operators (\ref{Sspher}).

Taking into account that initial reduced observables depend only on the certain phase space arguments,
we deduce the reduced representation of expansions (\ref{sedc}):
\begin{eqnarray}\label{rsedc}
   &&\hskip-12mm B(t)=\sum\limits_{n=0}^{\infty}\frac{1}{n!}\,\sum\limits_{k=0}^{n}\,(-1)^{n-k}\,
      \frac{n!}{k!(n-k)!}\,(\mathfrak{a}^{+})^{n-k}S(t)(\mathfrak{a}^{+})^{k}B(0)=\\
   &&S(t)B(0)+\sum\limits_{n=1}^{\infty}\frac{1}{n!}
      \big[\ldots\big[S(t),\underbrace{\mathfrak{a}^{+}\big],\ldots,\mathfrak{a}^{+}}_{\hbox{n-times}}\big]B(0)=\nonumber\\
   && e^{-\mathfrak{a}^{+}}S(t)e^{\mathfrak{a}^{+}}B(0).\nonumber
\end{eqnarray}
Therefore, in component-wise form, the generating operators of these expansions represented
as expansions (\ref{sed}) are the following reduced cumulants of groups of operators (\ref{Sspher}):
\begin{eqnarray}\label{rcc}
  &&\hskip-5mm U_{1+n}(t,\{1,\dots,s-n\},s-n+1,\dots,s)=
              \sum^n_{k=0}(-1)^{k}\frac{n!}{k!(n-k)!}S_{s-k}(t,1,\dots,s-k), \\
  &&\hskip-5mm n\geq 0.\nonumber
\end{eqnarray}

Indeed, solutions of the recursive relations (\ref{cexd}) with respect to first-order cumulants
can be represented as expansions in terms of cumulants acting on variables on which the initial
reduced observables depend, and in terms of cumulants not acting on these variables
\begin{eqnarray*}
   &&\hskip-5mm \mathfrak{A}_{1+n}(t,\{(1,\ldots,s)\setminus(j_1,\ldots,j_{n})\},j_1,\ldots,j_{n})=\\
   &&\sum_{Y\subset (j_1,\ldots,j_{n})}
       \mathfrak{A}_{1}(t,\{(1,\ldots,s)\setminus((j_1,\ldots,j_{n})\cup Y)\})
       \sum_{\mathrm{P}:\,(j_1,\ldots,j_{n})\setminus Y ={\bigcup\limits}_i X_i}
       (-1)^{|\mathrm{P}|}\,|\mathrm{P}|!\,\prod_{i=1}^{|\mathrm{P}|}\mathfrak{A}_{1}(t,\{X_{i}\}),
\end{eqnarray*}
where ${\sum\limits}_{Y\subset(j_1,\ldots,j_{n})}$ is the sum over all possible subsets
$Y\subset(j_1,\ldots,j_{n})$. Then taking into account the identity (\ref{aq}) and the equalities (\ref{id})
we derive expansions (\ref{rsedc}) over reduced cumulants (\ref{rcc}).

We note that traditionally, the solution of the BBGKY hierarchy for states of many hard spheres is represented
by the perturbative series \cite{PG90},\cite{GP90}. The expansions (\ref{rsedc}) can also be represented as
expansions (iterations) of perturbation theory \cite{BGer}:
\begin{eqnarray*}
   &&\hskip-12mm B(t)=\sum\limits_{n=0}^{\infty}\,\int\limits_{0}^{t} dt_{1}\ldots
        \int\limits_{0}^{t_{n-1}}dt_{n}S(t-t_{1})\big[\mathcal{L},\mathfrak{a}^+\big]
        S(t_1-t_2)\ldots S(t_{n-1}-t_n)\big[\mathcal{L},\mathfrak{a}^+\big]S(t_{n})B(0).\nonumber
\end{eqnarray*}
Indeed, as a result of applying analogs of the Duhamel equation to generating operators (\ref{dcumulant})
of expansions (\ref{sed}) we derive in component-wise form, for examples,
\begin{eqnarray*}
    &&\hskip-5mm U_{1}(t,\{1,\ldots,s\})=S_s(t,1,\ldots,s),\\
    &&\hskip-5mm U_{2}(t,\{(1,\ldots,s)\setminus(j_1)\},j_1)=\int\limits_{0}^{t}dt_{1}S_s(t-t_{1},1,\ldots,s)
      \sum_{j_{2}=1,\,j_2\neq j_{1}}^s\mathcal{L}_{\mathrm{int}}(j_1,j_{2})S_{s-1}(t_1,(1,\ldots,s)\setminus j_1),\\
   &&\vdots
\end{eqnarray*}

In Conclusion, we emphasize that within the framework of an equivalent approach to describing the evolution
of many hard spheres as the evolution of states (\ref{rdf}), an analog of the above results is valid. The
structure of the expansions, which are a non-perturbative solution to the Cauchy problem of the BBGKY
hierarchy for hard spheres, is considered in the next section.


\textcolor{blue!55!black}{\section{The BBGKY hierarchy for colliding particles}}

As mentioned above, the solution of the Cauchy problem of the BBGKY hierarchy for hard spheres
is traditionally represented as a series of the perturbation theory (an iteration series with
respect to the evolution of the state of a selected group of particles) \cite{B1},\cite{L75},\cite{GP85}.
This representation of the solution is applied to construct its Boltzmann-Grad asymptotics, which
is governed by the Boltzmann kinetic equation \cite{Sp91}--\cite{CGP97},\cite{PG90}. This section
considers methods for constructing a solution to the Cauchy problem for the BBGKY hierarchy of
evolution equations, including those not based on perturbation theory.


In this case, the sequence of reduced distribution functions is determined by the hierarchy of evolution
equations, known as the BBGKY hierarchy, whose generator is the operator adjoint to the generator of the
hierarchy of evolution equations for reduced observables in the sense of mean value functional.

It should be noted that for the system of a fixed, finite number of hard spheres, the BBGKY hierarchy
is an equation system for a finite sequence of reduced distribution functions. Such an equation system
is equivalent to the Liouville equation for the distribution function, which describes all possible states
of finitely many hard spheres. For a system of an infinite number of hard spheres, the BBGKY hierarchy is
an infinite chain of evolution equations, which can be derived as the thermodynamic limit of the BBGKY
hierarchy of a fixed finite number of hard spheres \cite{CGP97}.


For mean value functional (\ref{B(t)}) of observables the following representation holds:
\begin{eqnarray*}\label{F(t)}
   &&(B(t),F(0))=(B(0),F(t)),
\end{eqnarray*}
where a sequence of reduced distribution functions is defined as follows (known as the non-equilibrium
grand canonical ensemble \cite{PG83}):
\begin{eqnarray}\label{rdf}
   &&F(t)=(I,D(0))^{-1}e^{\mathfrak{a}}S^{\ast}(t)D(0).
\end{eqnarray}
On the space $L^{1}_{\alpha}=\oplus^{\infty}_{n=0}\alpha^n L^{1}_{n}$ of sequences of integrable functions,
the group of operators $S^\ast(t)=\oplus_{n=0}^\infty S^\ast_n(t)$ is an adjoint to the group of operators
(\ref{Sspher}) in the sense of mean value functional (\ref{averageD}) and is defined as follows \cite{PG90}:
\begin{eqnarray}\label{S*}
      && S^{\ast}(t)=S(-t).
\end{eqnarray}

In the space $L^{1}_{n}$ one-parameter mapping (\ref{S*}) is an isometric strong continuous
group of operators, i.e., it is a $C_{0}$-group.

The infinitesimal generator $\mathcal{L}=\oplus_{n=0}^\infty\mathcal{L}_{n}$ of the group of
operators (\ref{S*}) has the following structure:
\begin{eqnarray*}
 &&\mathcal{L}_{n}^\ast\doteq\sum_{j=1}^{n}\mathcal{L}^\ast(j)+
               \sum_{j_{1}<j_{2}=1}^{n}\mathcal{L}_{\mathrm{int}}^\ast(j_{1},j_{2}),
\end{eqnarray*}
and for $t>0$ the operators
$\mathcal{L}^{\ast}(j)$ and $\mathcal{L}_{\mathrm{int}}^\ast(j_{1},j_{2})$ are defined by the formulas:
\begin{eqnarray*}\label{L}
     &&\hskip-7mm \mathcal{L}^{\ast}(j) f_{n}\doteq-\langle p_j,\frac{\partial}{\partial q_j}\rangle f_{n},\\
\label{Lint}
     &&\hskip-7mm \mathcal{L}_{\mathrm{int}}^\ast(j_{1},j_{2})f_{n}
        \doteq\sigma^2\int\limits_{\mathbb{S}_{+}^2}d\eta\langle\eta,(p_{j_{1}}-p_{j_{2}})\rangle
        \big(f_n(x_1,\ldots,p_{j_{1}}^*,q_{j_{1}},\ldots,\\
     &&p_{j_{2}}^*,q_{j_{2}},\ldots,x_n)\delta(q_{j_{1}}-q_{j_{2}}+\sigma\eta)-
        f_n(x_1,\ldots,x_n)\delta(q_{j_{1}}-q_{j_{2}}-\sigma\eta)\big),\nonumber
\end{eqnarray*}
respectively, and notations accepted in formula (\ref{int}) are used, i.e., the pre-collision momenta
$p_{j_{1}}^*,p_{j_{2}}^*$ are defined by equalities (\ref{momenta}). For $t<0$ these operators on the
set $L^{1}_{n,0}$ are defined by the corresponding expressions \cite{CGP97}.

We note that the evolution of a state, i.e., the sequence of probability distribution functions
\begin{eqnarray*}
   &&D(t)=S^{\ast}(t)D(0),
\end{eqnarray*}
is a global-in-time solution to the Cauchy problem for a sequence of the weak formulation of the
Liouville equations for finitely many hard spheres.

The evolution of sequence (\ref{rdf}) of reduced distribution functions is governed by the Cauchy
problem of the BBGKY hierarchy for many hard spheres \cite{B1},\cite{CGP97},\cite{GP90}:
\begin{eqnarray}\label{h}
   &&\frac{d}{dt}F(t)=\mathcal{L}^\ast F(t)+\big[\mathfrak{a},\mathcal{L}^\ast\big]F(t),\\
   \nonumber\\
   \label{hi}
   &&F(t)|_{t=0}=F(0),
\end{eqnarray}
where the symbol $\big[\cdot,\cdot \big]$ denotes the commutator of operator (\ref{a}) and of the
Liouville operator $\mathcal{L}^\ast$, which is the generator of the isometric group of operators
(\ref{S*}). Hence, in evolution equation (\ref{h}), the second term of its generator has the following
component-wise form:
\begin{eqnarray*}
   &&\hskip-12mm (\big[\mathfrak{a},\mathcal{L}^*\big]f)_{s}(x_1,\ldots,x_s)=
       \sum_{i=1}^s\,\int\limits_{\mathbb{R}^{3}\times\mathbb{R}^{3}} dx_{s+1}\,
       \mathcal{L}^*_{\mathrm{int}}(i,s+1)f_{s+1}(t,x_1,\ldots,x_{s+1}),\quad s\geq1,
\end{eqnarray*}
and abbreviations accepted in formula (\ref{int}) are used.

\textcolor{blue!55!black}{\section{A non-perturbative solution of the BBGKY hierarchy}}

A non-perturbative solution of the Cauchy problem of the BBGKY hierarchy (\ref{h}),(\ref{hi}) is a sequence
of reduced distribution functions represented by the following expansions into series \cite{GG21},\cite{GerRS}:
\begin{eqnarray}\label{se}
    &&\hskip-7mm F_s(t,x_1,\ldots,x_s)=(e^{\mathfrak{a}}\mathfrak{A}^\ast(t)F(0))_s(x_1,\ldots,x_s)=\\
    &&\sum_{n=0}^\infty\frac{1}{n!}\int\limits_{(\mathbb{R}^{3}\times\mathbb{R}^{3})^{n}}
                 \mathfrak{A}_{1+n}^\ast(t,\{1,\ldots,s\},
                 s+1,\ldots,s+n)F_{s+n}^0(x_{1},\ldots,x_{s+n})dx_{s+1}\ldots dx_{s+n},\nonumber\\
    &&\hskip-7mm s\geq1,\nonumber
\end{eqnarray}
where the mappings $\mathfrak{A}^\ast_{1+n}(t),\,n\geq0,$ are the generating operators which are represented
by the cumulant expansions with respect to the group $S^\ast(t)=\oplus_{n=0}^\infty S^\ast_n(t)$ of operators
(\ref{S*}). The sequence of expansions into series (\ref{se}) is dual to the sequence of expansions  (\ref{sed})
in the sense of mean value functional (\ref{B(t)}).

Taking into account the definition of the $\star$-product (\ref{Product}), the cluster expansions of the groups
of operators (\ref{S*}) are represented by the mapping ${\mathbb E}\mathrm{xp}_{\star}$ in the form
\begin{eqnarray*}\label{ce}
    &&S^\ast(t)={\mathbb E}\mathrm{xp}_{\star}\,\mathfrak{A}^\ast(t).
\end{eqnarray*}
In component-wise form cluster expansions are represented by the following recursive relations:
\begin{eqnarray}\label{cex}
   &&\hskip-12mm S_{s+n}^\ast(t,1,\ldots,s,s+1,\ldots,s+n)=
        \sum\limits_{\mathrm{P}:\,(\{1,\ldots,s\},s+1,\ldots,s+n)=
       \bigcup_i X_i}\,\prod\limits_{X_i\subset\mathrm{P}}\mathfrak{A}^\ast_{|X_i|}(t,X_i),\quad n\geq 0,\nonumber
\end{eqnarray}
where the set consisting of one element of indices $(1,\ldots,s)$ we denoted by
the symbol $\{(1,\ldots,s)\}$ and the symbol ${\sum}_\mathrm{P}$ means the sum
over all possible partitions $\mathrm{P}$ of the set $(\{1,\ldots,s\},s+1,\ldots,s+n)$
into $|\mathrm{P}|$ nonempty mutually disjoint subsets $X_i$.

The solutions of recursive relations (\ref{cex}) are represented by the inverse mapping
${\mathbb L}\mathrm{n}_{\star}$ in the form of the cumulant expansion
\begin{eqnarray*}
    &&\hskip-8mm \mathfrak{A}^\ast(t)={\mathbb L}\mathrm{n}_{\star}(\mathbb{I}+S^\ast(t)).
\end{eqnarray*}
Then the $(1+n)th$-order cumulant of the groups of operators $S^\ast(t)=\oplus_{n=0}^\infty S^\ast_n(t)$
is defined by the following expansion:
\begin{eqnarray}\label{cumulant}
    &&\hskip-8mm \mathfrak{A}_{1+n}^\ast(t,\{1,\ldots,s\},s+1,\ldots,s+n)\doteq\\
    &&\sum\limits_{\mathrm{P}:\,(\{1,\ldots,s\},s+1,\ldots,s+n)={\bigcup}_i X_i}
       (-1)^{\mathrm{|P|}-1}({\mathrm{|P|}-1})!\prod_{X_i\subset\mathrm{P}}
       S^\ast_{|\theta(X_i)|}(t,\theta(X_i)),\nonumber
\end{eqnarray}
where the declusterization mapping $\theta$ is defined by the formula:
$\theta(\{1,\ldots,s\})=(1,\ldots,s)$ and the above notation is used.
The cumulants (\ref{cumulant}) of the groups of operators of the first two orders have the form:
\begin{eqnarray*}
    &&\hskip-8mm\mathfrak{A}^\ast_{1}(t,\{1,\ldots,s\})=S^\ast_{s}(t,1,\ldots,s),\\
    &&\hskip-8mm \mathfrak{A}^\ast_{1+1}(t,\{1,\ldots,s\},s+1)=
       S^\ast_{s+1}(t,1,\ldots,s+1)-S^\ast_{s}(t,1,\ldots,s)S^\ast_{1}(t,s+1),\\
    &&\vdots
\end{eqnarray*}

Note also that the first few terms of series (\ref{se}) were derived in papers \cite{G56},\cite{C62}
based on an analogue of cluster expansions of the reduced equilibrium distribution functions.

If $f_{s}\in L^{1}_{s}$, then taking into account that $\big\|S^\ast_{n}(t)\big\|_{L^{1}_{n}}=1$,
for the $(1+n)th$-order cumulant (\ref{cumulant}) the following estimate is valid:
\begin{eqnarray}\label{est}
   &&\hskip-12mm \big\|\mathfrak{A}^\ast_{1+n}(t)f_{s+n}\big\|_{L^{1}_{s+n}}
      \leq \sum\limits_{\mathrm{P}:\,(\{1,\ldots,s\},s+1,\ldots,s+n)={\bigcup}_i X_i}
      (|\mathrm{P}|-1)!\big\|f_{s+n}\big\|_{ L^{1}_{s+n}}\leq \\
   &&\sum\limits_{k=1}^{n+1}\mathrm{s}(n+1,k)(k-1)!\big\|f_{s+n}\big\|_{ L^{1}_{s+n}}
      \leq n!e^{n+2}\big\|f_{s+n}\big\|_{ L^{1}_{s+n}},\nonumber
\end{eqnarray}
where $\mathrm{s}(n+1,k)$ are the Stirling numbers of the second kind.

Then, according to this estimate (\ref{est}) for the generating operators of expansions (\ref{se}),
provided that the $\alpha>e$ series (\ref{se}) converges on the norm of the space $L^{1}_{\alpha}$,
and the inequality holds
\begin{eqnarray*}\label{Fes}
    &&\|F(t)\|_{L^{1}_{\alpha}}\leq c_{\alpha}\|F(0)\|_{L^{1}_{\alpha}},
\end{eqnarray*}
where $c_{\alpha}=e^{2}(1-\frac{e}{\alpha})^{-1}$. The parameter $\alpha$ can be interpreted as
the value inverse to the average number of hard spheres.

In fact, the following criterion holds.

\begin{criterion*}
A solution of the Cauchy problem of the BBGKY hierarchy (\ref{h}),(\ref{hi}) is represented by expansions
(\ref{se}) if and only if the generating operators of expansions (\ref{se}) are solutions of cluster
expansions (\ref{cex}) of the groups of operators (\ref{S*}).
\end{criterion*}

The necessary condition means that cluster expansions (\ref{cex}) are valid for groups of operators
(\ref{S*}). These recurrence relations are derived from definition (\ref{rdf}) of reduced
distribution functions, provided that they are represented as expansions (\ref{se}) for the solution
of the Cauchy problem of the BBGKY hierarchy (\ref{h}),(\ref{hi}).

The sufficient condition means that the infinitesimal generator of one-parameter mapping (\ref{se})
coincides with the generator of the BBGKY hierarchy (\ref{h}).
Indeed, in the space $L^{1}_{\alpha}$ the following existence theorem is true \cite{GR02}.
\begin{theorem*}
If $\alpha>e$, a non-perturbative solution of the Cauchy problem of the BBGKY hierarchy (\ref{h}),(\ref{hi})
is represented by series expansions (\ref{se}) in which the generating operators are cumulants of the
corresponding order (\ref{cumulant}) of groups of operators (\ref{S*}):
\begin{eqnarray}\label{sec}
    &&\hskip-18mm F_s(t,x_1,\ldots,x_s)=
         \sum_{n=0}^\infty\frac{1}{n!}\int\limits_{(\mathbb{R}^{3}\times\mathbb{R}^{3})^{n}}
         \,\sum\limits_{\mathrm{P}:\,(\{1,\ldots,s\},s+1,\ldots,s+n)={\bigcup}_i X_i}
         (-1)^{\mathrm{|P|}-1}({\mathrm{|P|}-1})!\times\\
    &&\hskip-7mm \prod_{X_i\subset\mathrm{P}}S^\ast_{|\theta(X_i)|}(t,\theta(X_i))
         F_{s+n}^0(x_{1},\ldots,x_{s+n})dx_{s+1}\ldots dx_{s+n}, \quad s\geq1.\nonumber
\end{eqnarray}
For initial data $F(0)\in L^{1}_{0}$ of finite sequences of infinitely differentiable functions
with compact supports sequence (\ref{sec}) is a unique global-in-time classical solution and for
arbitrary initial data $F(0)\in L^{1}_{\alpha}$ is a unique global-in-time generalized solution.
\end{theorem*}

The proof of this theorem is similar to the proof of the existence theorem for the Liouville equations
for many hard spheres in the space of sequences of integrable functions \cite{CGP97}.

%
\textcolor{blue!55!black}{\section{Representation of reduced distribution functions in the form of an iteration series}}
We observe that cluster expansions (\ref{cex}) of the groups of operators (\ref{S*}) underlie the
classification of possible solution representations (\ref{se}) of the Cauchy problem of the BBGKY
hierarchy (\ref{h}),(\ref{hi}). In a particular case, non-perturbative solution (\ref{sec})
of the BBGKY hierarchy for many hard spheres can be represented in the form of the perturbation
(iteration) series as a result of applying analogs of the Duhamel equation to cumulants
(\ref{cumulant}) of the groups of operators (\ref{S*}).

Indeed, let us put groups of operators in the expression of cumulant (\ref{cumulant}) into a new
order with respect to the groups of operators which act on the variables $(x_{1},\ldots,x_{s})$
\begin{eqnarray}\label{peregr}
    &&\hskip-21mm \mathfrak{A}^\ast_{1+n}(t,\{1,\ldots,s\},s+1,\ldots,s+n)=\\
    &&\hskip-21mm\sum\limits_{Y\subset\,(s+1,\ldots,s+n)}S^\ast_{s+|Y|}(t,(1,\ldots,s)\cup\,Y)
        \sum\limits_{\mathrm{P}\,:(s+1,\ldots,s+n)\setminus Y={\bigcup\limits}_i Y_i}
        (-1)^{|\mathrm{P}|}|\mathrm{P}|!\prod_{Y_i\subset\mathrm{P}}
        S^\ast_{|Y_{i}|}(t,Y_{i}).\nonumber
\end{eqnarray}
If $Y_{i}\subset(s+1,\ldots,s+n)$, then for the integrable functions $F_{s+n}^{0}$ and the unitary
group of operators $S^\ast(t)=\oplus_{n=0}^\infty S^\ast_n(t)$ the equality is valid
\begin{eqnarray*}
    &&\hskip-8mm\int\limits_{(\mathbb{R}^{3}\times\mathbb{R}^{3})^{n}}dx_{s+1}\ldots dx_{s+n}
      \prod_{Y_i\subset \mathrm{P}}S^\ast_{|Y_{i}|}(t;Y_{i})
        F_{s+n}^{0}(x_{1},\ldots,x_{s+n})=\\
    &&\int\limits_{(\mathbb{R}^{3}\times\mathbb{R}^{3})^{n}}dx_{s+1}\ldots dx_{s+n}F_{s+n}^{0}(x_{1},\ldots,x_{s+n}).
\end{eqnarray*}
Then, taking into account the equality
\begin{eqnarray*}
    &&\hskip-12mm\sum\limits_{\mathrm{P}\,:(s+1,\ldots,s+n)\setminus Y={\bigcup\limits}_i Y_i}
       (-1)^{|\mathrm{P}|}|\mathrm{P}|!=(-1)^{|(s+1,\ldots,s+n)\setminus Y|}, \quad Y\subset\,(s+1,\ldots,s+n),
\end{eqnarray*}
from expression (\ref{peregr}) for solution series expansions of the BBGKY hierarchy we obtain
\begin{eqnarray}\label{cherez1}
    &&\hskip-12mm F_s(t,x_1,\ldots,x_s)=\sum_{n=0}^\infty\frac{1}{n!}
       \int\limits_{(\mathbb{R}^{3}\times\mathbb{R}^{3})^{n}}U_{1+n}^\ast(t,\{1,\ldots,s\},\\
    &&\hskip+12mm s+1,\ldots,s+n)F_{s+n}^0(x_{1},\ldots,x_{s+n})dx_{s+1}\ldots dx_{s+n},\quad s\geq1,\nonumber
\end{eqnarray}
where $U_{1+n}^\ast(t)$ is the $(1+n)th$-order reduced cumulant of the groups of operators (\ref{S*})
\begin{eqnarray*}\label{rc}
    &&\hskip-12mm U_{1+n}^\ast(t,\{1,\ldots,s\},s+1,\ldots,s+n)=\\
    &&\hskip+5mm\sum\limits_{Y\subset (s+1,\ldots,s+n)}(-1)^{|(s+1,\ldots,s+n)\setminus Y|}
       S^\ast_{|(1,\ldots,s)\cup Y|}(t,(1,\ldots,s)\cup Y).\nonumber
\end{eqnarray*}
Using the symmetry property of initial reduced distribution functions, for integrand functions
in every term of series (\ref{cherez1}) the following equalities are valid
\begin{eqnarray*}
    &&\hskip-16mm\sum\limits_{Y\subset (s+1,\ldots,s+n)}
       (-1)^{|(s+1,\ldots,s+n)\setminus Y)|}S^\ast_{|(1,\ldots,s)\cup Y|}(t,(1,\ldots,s)\cup Y)
       F_{s+n}^0(x_{1},\ldots,x_{s+n})=\nonumber\\
    &&\hskip-8mm\sum\limits_{k=0}^{n}(-1)^{k}\sum\limits_{i_{1}<\ldots<i_{n-k}=s+1}^{s+n}
       S^\ast_{s+n-k}(t,1,\ldots,s,i_{1},\ldots,i_{n-k})F_{s+n}^0(x_{1},\ldots,x_{s+n})=\nonumber\\
    &&\hskip-8mm\sum\limits_{k=0}^{n}(-1)^{k}\frac{n!}{k!(n-k)!}
       S^\ast_{s+n-k}(t,1,\ldots,s+n-k)F_{s+n}^0(x_{1},\ldots,x_{s+n}).\nonumber
\end{eqnarray*}
Thus, the $(1+n)th$-order reduced cumulant represents by the expansion \cite{L61}:
\begin{eqnarray*}\label{Udef}
  &&\hskip-22mm U_{1+n}^\ast(t,\{1,\ldots,s\},s+1,\ldots,s+n)=
     \sum^n_{k=0}(-1)^k \frac{n!}{k!(n-k)!}S_{s+n-k}^\ast(t,1,\ldots,s+n-k),
\end{eqnarray*}
and consequently, we derive the representation for solution series expansions \cite{PG83} of
the BBGKY hierarchy which is written down in terms of the operator (\ref{a}):
\begin{eqnarray}\label{rcexp}
   &&\hskip-12mm F(t)= \sum\limits_{n=0}^{\infty}\frac{1}{n!}\,\sum\limits_{k=0}^{n}\,(-1)^{k}\,
      \frac{n!}{k!(n-k)!}\,\mathfrak{a}^{n-k}S^\ast(t)\mathfrak{a}^{k}F(0)=\\
   &&S^\ast(t)F(0)+\sum\limits_{n=1}^{\infty}\frac{1}{n!}
      \big[\underbrace{\mathfrak{a},\ldots,\big[{\mathfrak{a}}}_{\hbox{n-times}},S^\ast(t)\big]\ldots\big]F(0)=\nonumber\\
   && e^{\mathfrak{a}}S^\ast(t)e^{-\mathfrak{a}}F(0).\nonumber
\end{eqnarray}
Finally, in view of the validity of the equality
\begin{eqnarray*}
   &&S^\ast(t-\tau)\big[\mathfrak{a},\mathcal{L}^\ast\big]S^\ast(\tau)F(0)=
       \frac{d}{d\tau}S^\ast(t-\tau)\mathfrak{a}S^\ast(t(\tau)F(0),
\end{eqnarray*}
expansion (\ref{rcexp}) is represented in the form of perturbation (iteration) series of the BBGKY hierarchy
(\ref{h}) for hard spheres
\begin{eqnarray*}
   &&\hskip-12mm F(t)=\sum\limits_{n=0}^{\infty}\,\int\limits_{0}^{t} dt_{1}\ldots
        \int\limits_{0}^{t_{n-1}}dt_{n}S^\ast(t-t_{1})\big[\mathfrak{a},\mathcal{L}^\ast\big]
        S^\ast(t_1-t_2)\ldots S^\ast(t_{n-1}-t_n)\big[\mathfrak{a},\mathcal{L}^\ast\big]S^\ast(t_{n})F(0),\nonumber
\end{eqnarray*}
or in component-wise form \cite{L75},\cite{GP85}:
\begin{eqnarray*}\label{iter}
   &&\hskip-12mm F_s(t,x_1,\ldots,x_s)=\\
   &&\hskip-12mm\sum\limits_{n=0}^{\infty}\,\int\limits_{0}^{t}dt_{1}\ldots
       \int\limits_{0}^{t_{n-1}}dt_{n}\int\limits_{(\mathbb{R}^{3}\times\mathbb{R}^{3})^{n}}
       dx_{s+1}\ldots dx_{s+n}\,S^\ast_s(t-t_{1})
       \sum\limits_{j_1=1}^{s}\mathcal{L}^{\ast}_{\mathrm{int}}(j_1,s+1)\ldots
       \nonumber\\
   && \hskip-12mm S^\ast_{s+1}(t_1-t_2)\ldots S^\ast_{s+n-1}(t_{n-1}-t_n)
      \sum\limits_{j_n=1}^{s+n-1}\mathcal{L}^{\ast}_{\mathrm{int}}(j_n,s+n)
      S^\ast_{s+n}(t_{n})F_{s+n}^0(x_1,\ldots,x_{s+n}), \quad  s\geq1.\nonumber
\end{eqnarray*}
Some comments concerning the existence of the iteration series of the BBGKY hierarchy for hard spheres
were made above in Introduction.


\textcolor{blue!55!black}{\section{Conclusion}}

The structure of expansions (\ref{sed}) that represent a non-perturbative solution of the Cauchy
problem of the dual BBGKY hierarchy (\ref{dh}),(\ref{dhi}) for observables of many hard spheres
as well as expansions into series (\ref{se}) that represent a non-perturbative solution of the
BBGKY hierarchy (\ref{h}),(\ref{hi}) for their state was above justified.

Non-perturbative solutions to these hierarchies are represented in the form of expansions over
the particle clusters whose evolution is governed by the corresponding order cumulants
(\ref{dcumulant}) of the groups of operators (\ref{Sspher}) or cumulants (\ref{cumulant}) of
the groups of adjoint operators (\ref{S*}) for finitely many hard spheres. Cluster expansions
(\ref{cexd}) of the groups of operators (\ref{Sspher}) and cluster expansions (\ref{cex}) of
the groups of adjoint operators (\ref{S*}) underlie the classification of all possible solution
representations (\ref{sed}) and (\ref{se}) of the dual BBGKY hierarchy (\ref{dh}),(\ref{dhi})
and the BBGKY hierarchy (\ref{h}),(\ref{hi}), respectively.

Recall that the mean value functional (\ref{B(t)}) exists if $B(0)\in C_{\gamma}$ and
$F(0)\in L^{1}_{\alpha}$. In the case of the reduced observable of a number of hard spheres,
$N^{(1)}(t)=(0,1,0,\ldots)$, this means that
\begin{eqnarray*}
    &&\big|(N^{(1)}(t),F(0))\big|=\big|\int\limits_{\mathbb{R}^{3}\times\mathbb{R}^{3}}\,
       F_{1}^0(x_{1})dx_{1}\big|\leq\big\|F_{1}^0\big\|_{L^{1}_{1}}<\infty.
\end{eqnarray*}
Consequently, the states of a finite number of hard spheres are described by sequences of functions
from the space $L^{1}_{\alpha}$. To describe an infinite number of hard spheres, it is necessary to
consider reduced distribution functions from appropriate function spaces, for example, from the space
of sequences of bounded functions with respect to the configuration variables \cite{L75},\cite{GP85}.

Nowadays, in articles \cite{GG22},\cite{GG22R} an approach to the description of the evolution of
correlations for many hard spheres is developed that is based on a hierarchy of evolution equations
for the cumulants of the probability distribution functions governed by the Liouville equations. The
constructed dynamics of correlations underlie the description of the evolution of the states of many
hard spheres described by the BBGKY hierarchy for reduced distribution functions or the hierarchy of
nonlinear evolution equations for reduced correlation functions \cite{GG22}.

We point out that an approach to the description of the kinetic evolution of many hard spheres
within the framework of the evolution of observables was developed in papers \cite{GG18},\cite{G13}.
One of the advantages of such an approach to the derivation of the Boltzmann equation is an
opportunity to describe the process of the propagation of initial correlations in the low-density
scaling limit \cite{GG23},\cite{GG18},\cite{GG15}.

Recently, in papers \cite{GG11},\cite{GG12}, by means of the so-called kinetic cluster expansions
of cumulants of the groups of operators of finitely many hard spheres, the generalized Enskog kinetic
equation was rigorously derived (see also the survey \cite{GG21}). This approach has also been applied
to the derivation of kinetic equations in the case of many hard spheres with inelastic collisions
\cite{GB14}, as well as a generalization of the Fokker--Planck kinetic equation for open systems with
collisional dynamics was justified \cite{GG15},\cite{GG14}.

Thus, the concept of cumulants of the groups of operators underlies non-perturbative expansions of
solutions to hierarchies of fundamental equations that describe the evolution of observables and
of a state of many colliding particles \cite{GG21}, as well as the basis of the kinetic description
of the collective behavior of infinitely many colliding particles \cite{GG12}.

\bigskip
\textbf{Acknowledgements.} \,\,{\large\textcolor{blue!55!black}{Glory to Ukra\"{\i}ne!}}

\bigskip


\addcontentsline{toc}{section}{\textcolor{blue!55!black}{References}}

\vskip+5mm

\end{document}